# Static and Dynamic Signatures of Anisotropic Electronic Phase Separation in La$_{2/3}$Ca$_{1/3}$MnO$_3$ Thin Films under Anisotropic Strain

L. Hu, L.Q. Yu, P. Xiong

*Department of Physics, Florida State University, Tallahassee, Florida 32306, USA*

X.L. Wang, J.H. Zhao

*Institute of Semiconductors, Chinese Academy of Sciences, Beijing 100083, China*

L.F. Wang, Z. Huang, W.B. Wu

*Hefei National Laboratory for Physical Sciences at Microscale, University of Science and Technology of China, Hefei, Anhui 230026, China*

## ABSTRACT

The electronic phase separation (EPS) of optimally doped La$_{2/3}$Ca$_{1/3}$MnO$_3$ (LCMO) thin films under various degrees of anisotropic strain is investigated by static magnetotransport and dynamic relaxation measurements. Three LCMO films were grown simultaneously on (001) NdGaO$_3$ (NGO) substrates by pulsed laser deposition, and then post-growth annealed at 780 ºC in O$_2$ for different durations of time. With increasing annealing time, the films developed significant strains of *opposite signs* along the two orthogonal in-plane directions. The static temperature-dependent resistivity, ρ(T), was measured simultaneously along the two orthogonal directions. With increasing annealing time, both zero-field-cooled and field-cooled ρ(T) show significant increases, suggesting strain-triggered EPS and appearance of antiferromagnetic insulating (AFI) phases in a ferromagnetic metallic (FMM) ground state. Meanwhile, ρ(T) along the tensile-strained [010] direction becomes progressively larger than that along the compressive-strained [100]. The enhanced resistivity anisotropy indicates that the EPS is characterized by phase-separated FMM entities with a preferred orientation along [100], possibly due to the cooperative deformation and rotation/tilting of the MnO$_6$ octahedra under the enhanced anisotropic strain. The anisotropic EPS can also be tuned by an external magnetic field. During a field-cycle at several fixed temperatures, the AFI phases are melted at high fields and recovered at low fields, resulting in sharp resistance changes of ratio as high as 10$^4$. Furthermore, the resistivity was found to exhibit glass-like behavior, relaxing logarithmically in the phase-separated states. Fitting the data to a phenomenological model, the resulting resistive viscosity and characteristic relaxation time are found to evolve with temperature, showing a close correlation with the static measurements in the EPS states.

## I. INTRODUCTION

In the mixed-valence manganites, the electronic phase separation (EPS) is considered central to a host of fascinating properties such as the metal-insulator transition (MIT) and colossal magnetoresistance (CMR).[1-4] EPS refers to the spatial coexistence of phases with different electronic and/or magnetic properties in the absence of compositional variations (chemical segregation).[1,2,5] These striking phenomena arise from the competition and coupling of the charge, spin, orbital and lattice degrees of freedom.[2,5] Within a network of corner-sharing $MnO_6$ octahedra, slight variations in length and angle of the Mn-O bonds, or rotation and tilt of the octahedra, can make large differences in electronic and magnetic properties.[6] Taking the instance of chemical doping, cations of different valences and sizes can not only change the charge carrier density and the size of an octahedron, but also induce Jahn-Teller (JT) distortion and modify the double exchange/super exchange interactions,[7,8] engendering rich evolving phases. The induced MIT is commonly accompanied by CMR, which has been intensively investigated. In the prototypical $La_{0.75}Ca_{0.25}MnO_3$ films, the observation of electrical nonlinearity near the MIT strongly suggests the existence of correlated polarons.[9] Near certain phase boundaries in the strongly correlated materials, coexistence of multiple phases has been observed.[1] These spatially inhomogeneous phases with various length scales suggest complex underlying mechanisms for the EPS. In $La_{1-x}Ca_xMnO_3$, at certain doping levels ($x > 0.5$), the carriers tend to form periodical stripes at nanometer scale with corresponding lattice distortions, which has been observed through high-resolution imaging by transmission electron microscopy.[10,11] The origin of the stripe formation is suggested to be triggered by the minimization of the inherent strain energy from local chemical disorder[12,13] and cooperative lattice distortion.[14] Another larger scale EPS with coexisting ferromagnetic metallic (FMM) and charge ordered insulating (COI) phases has been extensively investigated in the compound $La_{(5/8-y)}Pr_yCa_{3/8}MnO_3$ (LPCMO).[3] The mechanism of this large (micrometer) scale electronic inhomogeneity appears to be beyond the charge ordering at atomic scale, since the electronic charge density at long length scales over a few lattice constants should be nearly constant due to long-range Coulomb interaction. Moreover, local chemical inhomogeneity alone cannot satisfactorily explain the large-scale phase coexistence; X-ray diffraction analysis indicated no chemical variation within the two phases.[3] On the other hand, elastic strain stemming from cooperative lattice distortion is known to be crucial for the self-organized EPS at large scales.[3,4,15-17] Many theoretical models have explicitly invoked the essential role of strain in EPS.[16,18,19] Primarily relying on the structural aspect, these theories have emphasized that the electron-lattice interaction accompanying the coupling between short and long range strains can modify the intrinsic elastic energy landscape and give rise to a self-organized phase separation over both nanometer- and micrometer-scales.[16] On top of that, experimentally, long-range anisotropic elastic strain induced in LPCMO thin films from lattice mismatch with the substrates have given rise to a remarkable anisotropic distribution of coexisting phases in EPS.[20] As a consequence of the substrate-induced anisotropic tensile strains, the MIT in these systems exhibits substantial anisotropies manifested in the resistivity and transition temperatures along the anisotropically strained directions.[20]

In fact, thin films with epitaxial strain have been an effective platform for investigating the structural effects of octahedral configurations on EPS in manganites. Distinct from the bulk materials, fine control and deterministic tuning of the strain are achieved in epitaxial thin films by interfacial engineering between the films and substrates and in multilayer superlattices with different lattice structures.[21] Through epitaxial growth, different choices of substrate and interlayer in superlattices have been shown to result in significantly different physical properties for films of identical chemical compositions.[9,22,23]

As another notable feature of the EPS, glassy behavior of the EPS state has been frequently observed in manganites, manifested in time relaxation of magnetization and resistivity.[2,24] The dynamics of the EPS is known to be influenced by the competition between coexisting phases. Due to comparable free energies of the competing phases in the EPS state, these phases switch into each other under external stimuli, often accompanied by structural changes.[25,26] The commonly observed relaxation in manganites has a logarithmic time dependence. Logarithmic relaxation of the resistance was observed by Helmolt *et al* in La$_{2/3}$Ca$_{1/3}$MnO$_{3+\delta}$ thin films,[27] which is expected for a flat distribution of energy barriers separating various magnetic phases.[28] Fisher *et al.* tracked the correlation between the relaxations of resistivity and magnetization, and further qualitatively explained the dynamics within a framework of distributed energy barriers.[29] Additional exponential-like time relaxations were also observed,[29] implying other factors associated with energy barrier distribution,[30] such as structural defects and strain. Of particular relevance here, strain in thin films could potentially alter the energy barriers and impact the spatial phase distributions in the EPS state.[16]

In this paper, we report an investigation of the effects of anisotropic strain on the EPS in optimally doped LCMO films epitaxially grown on (001) NGO, via a combination of static and dynamic resistivity measurements with varying temperature and magnetic field along the two orthogonal in-plane directions. The strain in the films was created by post-growth annealing and controlled by the annealing time. In samples subject to prolonged annealing, $\rho(T)$ is found to depend sensitively on the thermal and magnetic field histories, which exhibits distinct behaviors after ZFC and FC in the intermediate temperature ranges, implying different emergent phases and EPS states. The simultaneous static transport measurements along the two in-plane directions reveal significant resistivity anisotropies in the phase-separated states tuned by temperature and field, which is enhanced with increasing annealing time. These pronounced anisotropies in the EPS state are consistent with an orientation-preferred growth of the FMM phase in an AFI background,[31,32] possibly due to the enhanced structural deformation and rotation/tilt of the MnO$_6$ octahedra under the anisotropic strain.[21] The dynamic relaxations of the resistivity, $\rho(t)$, in the EPS state show logarithmic time dependence. Analyzed by a phenomenological model, the resistivity relaxations are found to proliferate with increasing annealing time, and their evolution with temperature correlates closely with the signatures of the EPS in the static measurements.

## II. EXPERIMENT

Three 48-nm thick films of optimally doped La$_{2/3}$Ca$_{1/3}$MnO$_3$ were grown on (001) NGO substrates by pulsed laser deposition under identical conditions, at substrate temperature of 735°C and oxygen pressure of 45 Pa. The films were subsequently annealed *ex situ* at 780°C in an oxygen atmosphere for 1.5, 5 and 20 hours. The room-temperature *Pbnm* lattice parameters, octahedral rotation angle about [001], and tilt angle about [100] for bulk LCMO (NGO) at room temperature are: $a$ = 5.4717 Å (5.4332 Å), $b$ = 5.4569 Å (5.5034 Å) and $c$ = 7.7112 Å (7.7155 Å), $\varepsilon$ = 6.36° (9.09°), and $\delta$ = 9.57° (13.45°), respectively.[33,34] The annealing promotes film-substrate coherency and thus induces anisotropic strain, accompanied by increased rotation and tilt of the MnO$_6$ octahedra in the film. A fully strained LCMO/(001)NGO film thus has -0.70% compressive strain along [100] and +0.85% tensile strain along [010], as depicted in Fig. 1(a)).

In order to examine the effects of the anisotropic strain on the magnetoelectronic states and transport, all samples were patterned into identical "L-bar" devices by standard photolithography and wet-etching in a HPO$_3$ (3.4%) and H$_2$O$_2$ (1.5%) acid solution. As shown in the schematic diagram of the device in Fig. 1(b), one arm of the "L-bar" is aligned along the compressive-strained [100] direction and the other points in the tensile-strained [010] direction. Constant DC currents ranging from 0.5 µA to 0.5 mA were applied across the "L-bar" for simultaneous transport and relaxation measurements in the two orthogonal directions. An I-V curve was taken before or after each measurement to ensure the measured voltage has a linear dependence on the applied current, and Joule heating is negligible. The static magnetotransport measurements were conducted in a custom $^4$He cryostat, and the dynamic relaxation measurement was carried out in a Quantum Design physical property measurement system (PPMS) for ease of automated control of time sequences.

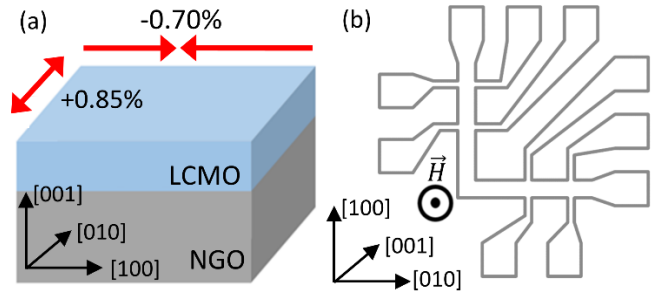

**Fig. 1.** (a) Schematic diagrams depicting the structure and crystalline directions for the LCMO/NGO films; the arrows indicate the compressive and tensile strain along [100] and [010] respectively. (b) A schematic diagram of an "L-bar device" with the two arms pointing along [100] and [010]. The magnetic field is applied perpendicular to the films in the magnetotransport measurements.

### III. RESULTS
#### A. Static anisotropic transport: temperature dependence

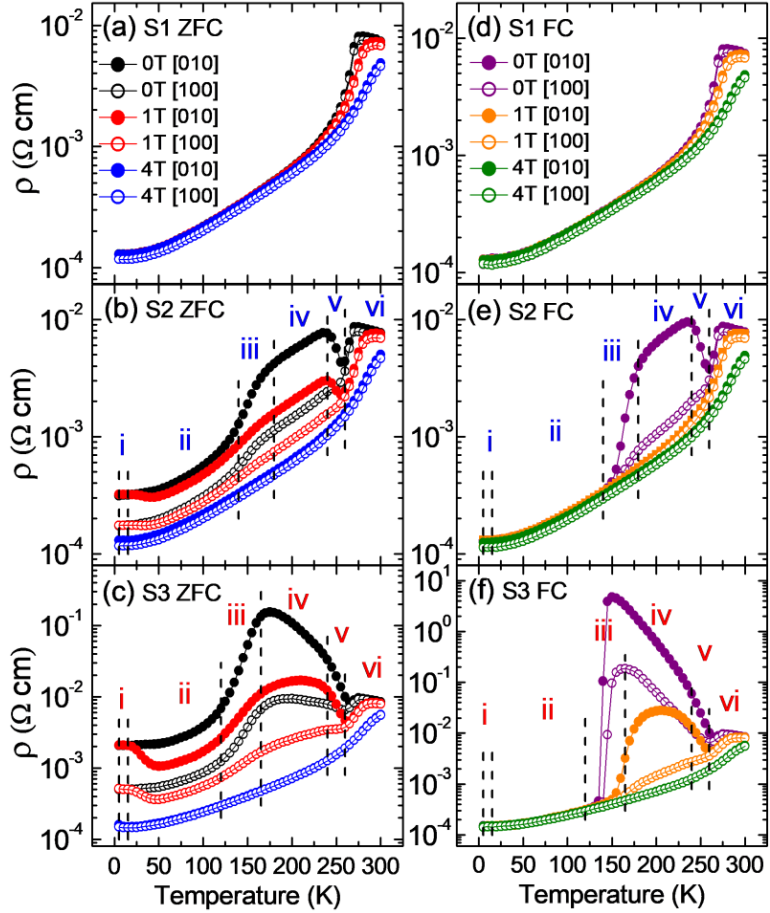

**Fig. 2.** (a) - (c) Temperature dependent resistivities, $\rho(T)$, along [100] and [010] directions in different perpendicular magnetic fields after zero-field-cooling (ZFC) for the three samples. (d) - (f) $\rho(T)$ data after field-cooling (FC) (4 T) for the same three films. All measurements were taken during warming up.

Fig. 2 shows the resistivities along the two in-plane directions as functions of temperature for the three samples measured while being warmed up after zero-field-cooling (ZFC, Figs. 2(a) – 2(c)) and field-cooling (FC, Figs. 2(d) – 2(f)) at 4 T to the base temperature of 5 K. For Sample 1 (S1), which was annealed briefly for 1.5 h, $\rho(T)$ shows essentially bulk-like behavior, with a paramagnetic insulating (PMI) to FMM phase transition and concomitant CMR effect around 270 K. Across this MIT, the differences of the resistivities between the two directions are negligible (< 0.5 m$\Omega$cm), indicating no significant anisotropy. Also, there is little difference between the ZFC and FC $\rho(T)$. Therefore, *S1 serves as a good reference sample for the bulk behavior of optimally doped LCMO*. With increasing annealing time, a number of qualitatively different features in $\rho(T)$ emerge in samples S2 (5 h) and become more pronounced in S3 (20 h). We first focus on the ZFC data (Figs. 2(b) and 2(c)) at different temperature ranges: i) Starting from the lowest temperature, in a range between 5 K and 15 K, $\rho$ along both [100] and [010] directions are essentially independent of temperature and magnetic field (up to 1 T). Similar low-temperature and low-field plateaus were previously observed in LPCMO,[35,36] considered as the prototype material exhibiting micrometer scale EPS, which are ascribed to a frozen mixed state of FMM and AFI phases. ii) Between 15 K and 140 K (120 K) for S2 (S3), ρ in 0 T increases gradually with increasing temperature, while ρ(T) in 1 T of applied field shows an initial drop, followed by a gradual increase similar to that of ρ in 0 T. iii) In a narrow temperature range near 150 K (130 K) for S2 (S3), ρ in zero field shows a sharp increase with temperature. iv) The sharp resistivity increase is followed by a gradual increase (decrease) for S2 (S3) at a higher temperature 180 K (165 K). v) Between about 240 K and 260 K, the resistivity decreases more sharply, signaling a transition to a metallic state. vi) Above 260 K, both S2 and S3 revert to the bulk-like behavior, similar to that in S1, with a MIT at 270 K. An applied field of 4 T fully restores bulk-like ρ(T) for the entire measurement temperature range; all three samples show the same behavior.

These features in the temperature dependence of the resistivity clearly emerge from extended annealing of the samples, and depend sensitively on the magnetic field/thermal history. FC in a field of 4 T from 300 K to 5 K, has negligible effect on S1 but significantly alters the ρ(T) for both S2 and S3 at low temperatures, as shown in Figs. 2(d) - 2(f). First, at low temperatures up to ~120 K (140 K) in S2 (S3) (regimes i and ii above), ρ(T) become essentially field-independent and attain the same values for S1. Secondly, the rapid zero-field resistivity jumps at ~130 K become even sharper, especially for S3 in which the peak resistivities along both in-plane directions are enhanced by more than an order of magnitude.

We now focus on the emergent resistivity anisotropies in S2 and S3. One overarching feature is the close correlation between the presence of any resistivity anisotropy and deviation from the bulk behavior (as represented by S1). This correlation is strong evidence that the resistivity anisotropy is a consequence of the micrometer scale EPS, which is not present in bulk optimally doped LCMO. In Fig. 3, we plot the ratio of the zero-field resistivities along [010] and [100] after ZFC (open symbols) and FC (solid symbols) for the three samples, which provides a direct and quantitative visualization of resistivity anisotropy (RA(T)), defined as $\rho(T)_{[010]}/\rho(T)_{[100]}$, in different temperature ranges.

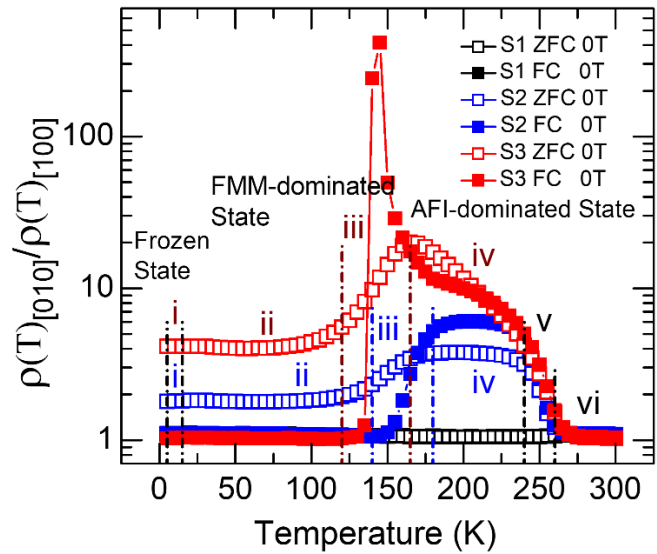

**Fig. 3.** Temperature dependent zero-field resistivity anisotropy (RA(T)), $\rho(T)_{[010]}/\rho(T)_{[100]}$, for the three samples after ZFC or FC (4 T). The EPS below 260 K is manifested in three distinct states, the frozen state, FMM-dominated state, and AFI-dominated state.

The RA is unity for S1 in the entire temperature range, while for S2 and S3, whenever the sample's resistivity attains the values for S1, the RA(T) also becomes unity. For FC, this includes the low temperature (regimes i and ii) and high temperature (regime vi) ranges; in between (regime iv) the RA(T) shows much enhanced values which are independent of (S2) or weakly dependent on (S3) temperature. On either side, the RA(T) decreases to unity in a well-defined transition. The lower temperature transition in regime iii is particularly striking, which for S3 takes the form of a sharp peak of more than 500-fold increase for the RA. For ZFC, the RA(T) in the intermediate temperature range (regimes iii, iv and v) are qualitatively similar, albeit the absolute values are much diminished. Moreover, although the ZFC RA(T) remain approximately temperature-independent at low temperatures (regimes i and ii), they are elevated above unity (1.8 for S2 and 4.2 for S3). It is interesting to note that although the R(T) data in Fig. 2 clearly reveal a frozen state below 20 K for both S2 and S3, the RA are virtually indistinguishable below and above 20 K. Based on both the static temperature dependencies of resistivity $\rho(T)$ and zero-field resistivity anisotropy RA(T), the EPS can be divided into a frozen state (regime i), FMM-dominated state (regime ii) and AFI-dominated state (regime iii, iv and v), which is consistent with previous studies.[37]

### B. Static anisotropic transport: magnetic field dependence

The temperature dependences of the resistivity indicate that the strained LCMO films present a system of complex phase configurations which evolve with temperature and magnetic field. The magnetic and electronic properties vary greatly with different thermal or magnetic field cycling.[38] To directly examine the evolution of various EPS states with the magnetic field, magnetoresistance (MR), $\rho(H)$, has been measured at several temperatures for S2 and S3 along the two in-plane directions. Fig. 4 presents the MR at six selected temperatures for S3, covering all the distinct states in the EPS; the MR for S2 is shown for two selected temperatures in regimes iii and iv, respectively.

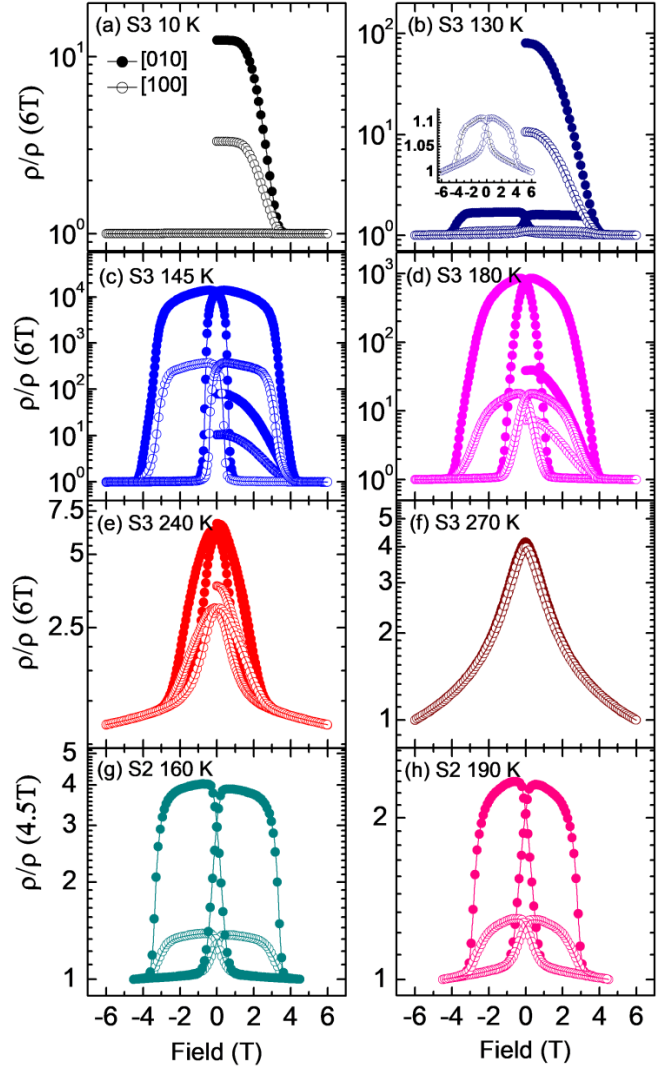

**Fig. 4**. The normalized field-dependent resistivity ρ(H) of S3 and S2 at selected temperatures (S3: (a) 10 K; (b) 130 K; (c) 145 K; (d) 180 K; (e) 240 K and (f) 270 K. S2: (g) 160 K and (h) 190 K) in both [010] and [100] directions. (The inset in (b): close-up view of the MR hysteresis along [100] direction.)

To ensure consistency in the MR measurement, each curve was taken in an identical sequence: After ZFC from 300 K to a target temperature, isothermal ρ(H) was measured with a low sweeping rate of 0.09 T/min from 0 T to a high magnetic field (4.5 T for S2, 6.0 T for S3), followed by a full field sweep cycle. At the highest fields, the resistivities for S2 and S3 are close to that for S1 over the entire temperature range, and there is no discernible difference between the resistivities along [010] and [100] directions. In Fig. 4, ρ(H) is normalized by its value at the highest measurement field to better reveal the anisotropy at low fields.

First, we focus on S3 (Figs. 4(a) – 4(f)). At 10 K in the frozen state, ρ(H)/ρ(6T) shows a plateau at low fields, indicating no melting of the AFI phase up to 1.4 T. Above 1.4 T the MR ratio drops steeply and approaches unity at 3.7 T, suggesting a complete field-induced melting of the AFI phase. In the subsequent field-sweep cycle, ρ(H) stays essentially constant, consistent with the FC ρ(T) results in Fig. 2(f). At 130 K as one approaches the boundary between the FMM-dominated state and the AFI-dominated state, a finite hysteretic MR appears for both in-plane directions in the field-sweep cycle after the initial up-sweep, suggesting re-entrance of the AFI phase and EPS at low fields. The most dramatic MR was obtained at 145 K, near the point of percolation indicated by the ρ(T) data. The field sweep reveals a much enhanced MR and resistivity anisotropy, with MR ratio (ρ(H)/ρ(6T)) of more than $10^4$ at low fields. At higher temperatures (180 K and 240 K), the MR are qualitatively similar, but the magnitude of MR ratio decreases precipitously and the MR switching become increasingly gradual. At 270 K, above the onset of the AFI phases, the hysteresis disappears and the MR reverts to bulk-like behavior. S2, with weaker anisotropic strains, exhibits a similar trend of MR, albeit with much reduced magnitude, as exemplified by the data at two selected temperatures in the AFI-dominated state shown in Figs. 4(g) and 4(h).

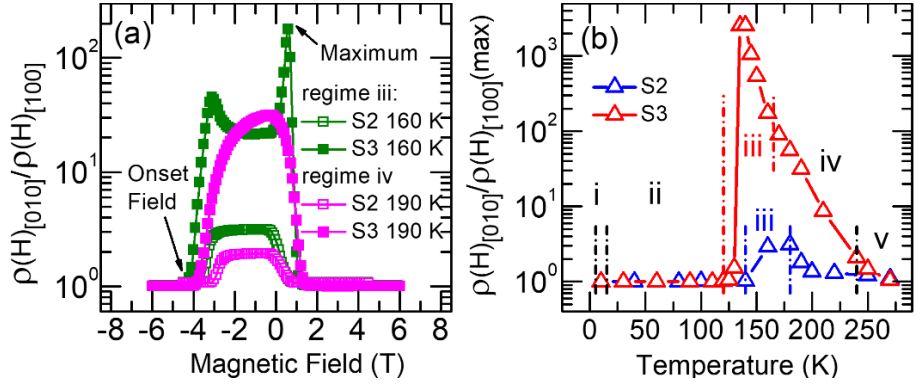

**Fig. 5**. (a) Typical field-dependent RA(H), $\rho(H)_{[010]}/\rho(H)_{[100]}$, curves for S2 and S3 at 160 K (regime iii) and 190 K (regime iv) in the AFI-dominated states. (b) The maximum values for all isothermal RA(H) curves measured at different temperatures for S2 and S3.

Based upon the results of the MR along the two directions, the field-dependent resistive anisotropy, RA(H), defined as $\rho(H)_{[010]}/\rho(H)_{[100]}$, can be evaluated. To quantitatively visualize RA(H) in the presence of EPS, isothermal $\rho(H)_{[010]}/\rho(H)_{[100]}$ for S2 and S3 are calculated from the MR measurements, and several representative RA(H) curves within regimes iii and iv in the AFI-dominated states for S2 and S3 are shown in Fig. 5(a). At 160 K (regime iii) in S3, RA(H) displays two peaks at 0.56 T and -3.07 T, and a plateau between those fields. This unique RA(H) is only present at temperatures from 140 K to 160 K in S3, and absent in S2. In S2 at the same temperatures, RA(H) exhibits a broad plateau of about 3.1 from -0.27 T to -2.91 T without any peaks. At 190 K (regime iv), RA(H) in S3 no longer has any peaks, while that in S2 maintains a plateau at a suppressed magnitude. In Fig. 5(b) we plot the *maximum* values of RA(H), $(\rho(H)_{[010]}/\rho(H)_{[100]})_{max}$, extracted from all the MR curves measured for S2 and S3, in order to directly compare the magnitude of RA(H) with varying magnetic field. It is evident that significant resistivity anisotropy, indicated by RA(H) of elevated values above unity, is present only in the AFI-dominated EPS states for S2 and S3. For S3 particularly, with increasing temperature from the FMM-dominated state to the AFI-dominated state, RA(H) increases sharply by over three orders of magnitude and peaks at 140 K, then it diminishes with increasing temperature and disappears at the onset temperature (260 K) for the AFI phase. For S2, RA(H) shows a similar trend but with much smaller magnitudes.

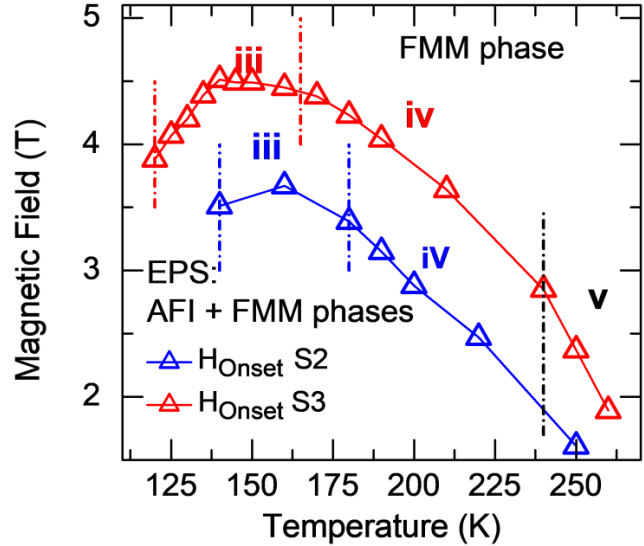

**Fig. 6**. H-T phase diagram with the field separating the FMM phase and the EPS state with AFI phase for S2 and S3.

Based on the MR data and the resulting RA(H), an H-T phase diagram of the EPS state and FMM phases for S2 and S3 is constructed, as shown in Fig. 6. We surmise that within a field sweep cycle, RA(H) values above unity corresponds to the reappearance of the AFI phase; when the applied field is adequate to suppresses the formation of the AFI phase, RA(H) drops to unity as that in S1 which is free of any AFI phase. We chose the magnetic field at which RA(H) reaches 1.01 as the point separating the two states. For both samples, the onset fields tend to have a similar temperature dependence as the maximum value of RA(H): With decreasing temperature, the onset field appears with the initial appearance of the AFI phase near the boundary of the FMM-dominated and AFI-dominated states, gradually reaches a peak, then decreases slightly with decreasing temperature, eventually disappears as no AFI phase re-enters in the field sweep cycle. The onset field also becomes significantly higher with increasing annealing time for S3.

### C. Temperature-dependent resistivity relaxation

The EPS is also manifested in the dynamic behavior of the resistivity, which exhibits dependence closely correlated with the static signatures of the EPS. The resistivity relaxation was measured over extended times at various temperatures following an identical FC sequence: After FC (4 T) to 5 K, the sample was warmed in zero field to a target temperature. Care was taken so that the target temperature was reached without significant temperature overshoot. After the temperature was stabilized (±5 mK), $\rho(t)$ was measured at the fixed temperature for $10^4$ s. A full set of results for S3 is shown in Fig. 7((a)-(h)); for S2, the relaxation measurements were performed for several selected temperatures, as shown in Fig. 7((i)-(l)). Each resistivity curve was normalized with respect to its initial (t = 0) value $\rho_0$.

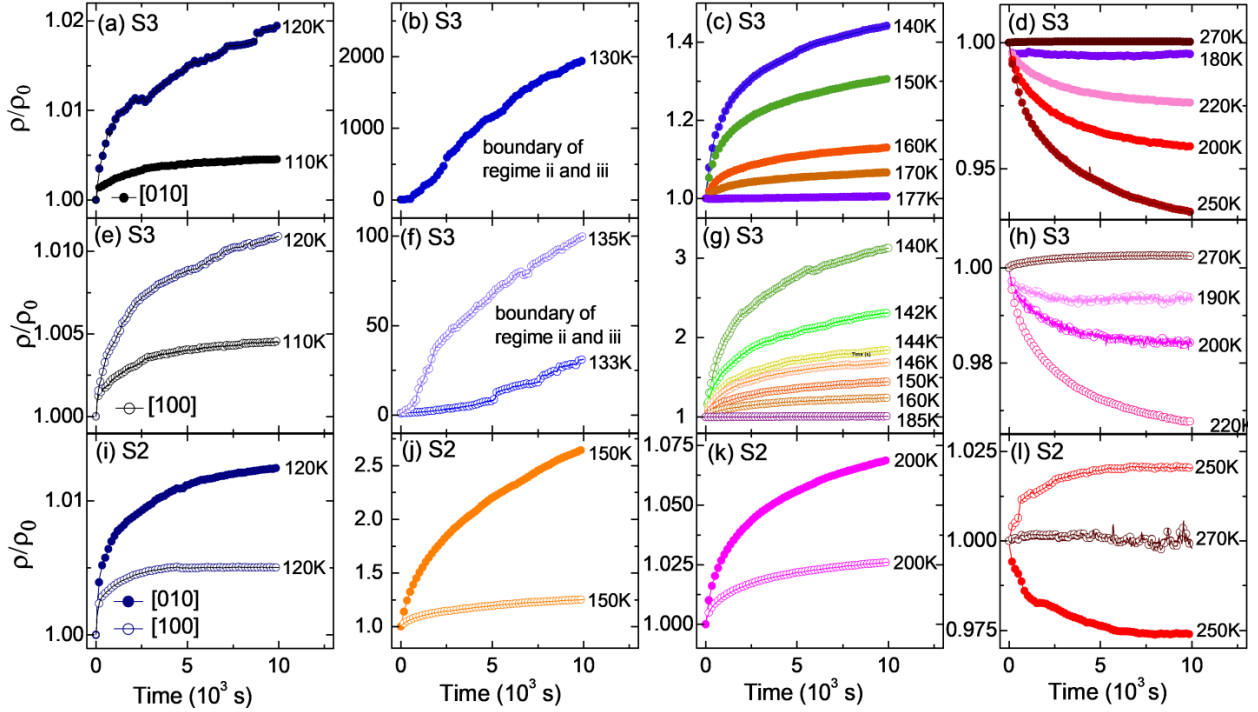

**Fig. 7.** Time dependence of the normalized resistivity, $\rho(t)/\rho(t = 0)$, along [010] ((a) – (d)) and [100] ((e) – (h)) directions for S3. $\rho(t)/\rho(t = 0)$ data along [010] and [100] directions at selected temperatures (120 K, 150 K, 200 K, 250 K and 270 K) for S2.

Focusing on S3, at 110 K and 120 K in the regime of percolated FMM phase, no substantial resistivity change with time was observed in either direction (<1.3% after $10^4$ s). As the temperature approaches the percolation point of the FMM phase in the AFI background, strikingly large resistivity variation appears. At first, $\rho$ grows by up to 2000 times (at 130 K) and 100 times (135 K) of its respective initial value at the end of measurement along [010] and [100] respectively. The rise of resistivity indicates significant growth of the AFI phase with time. In addition, $\rho(t)/\rho_0$ exhibits a multitude of burst-like jumps, reflecting the avalanche-like growth of the AFI phase. Similar avalanche-like dynamics was reported in another phase-separated manganese oxide ($Pr_{0.6}Ca_{0.4}Mn_{0.96}Ga_{0.04}O_3$) in a field-driven transformation.[39] Into the AFI-dominated region, the magnitude of the positive relaxation (increase of resistivity) decreases sharply but remains significant up to about 177 K and 185 K for [010] and [100] directions respectively. Beyond these temperatures *the resistivity relaxations change sign and become negative*. These observations suggest that there are two distinct dynamic states in the AFI-dominated regime: At low temperatures, the energetics favor the growth of the AFI domains, at increasingly slower rate with increasing temperature; beyond a well-defined temperature and with increasing temperature, it becomes more energetically favorable for the FMM domains to grow. At 177 K for [010] and 185 K for [100], the resistivity remains essentially constant over the entire period of measurement, indicating *a state of dynamic equilibrium* near these temperatures. Above the onset temperature of the AFI (~260 K) domains, the resistivity relaxation disappears. For S2, the resistivity relaxation was measured at a limited number of temperatures, and the results are shown in Fig. 7((i)-(l)). The general trend of the resistivity relaxation is similar to that in S3, but with some notable differences: near the boundary of the FMM-dominated and AFI-dominated states (150 K for S2), the resistivity relaxations along both directions show much smaller magnitudes than those in S3, and the burst-like jumps are absent. With increasing temperature, the resistivity relaxation along [010] turns negative around 250 K, while it remains positive along [100], suggesting distinct dynamics of the AFI domains along the two directions and absence of a dynamic equilibrium state in EPS in S2.

The dynamic relaxations in perovskite manganites bear much resemblance to the magnetization dynamics in spin glasses and hard magnets in many aspects.[40,41] The magnetic relaxation in these systems was considered to be a thermal activation process following a sudden change of the applied magnetic field. The process achieves the thermally-assisted magnetization switching over the energy barriers of different magnetic configurations.[42] In most cases, the observed magnetization relaxations have a logarithmic time dependence,[43] which can be described by $M(t) = M_0 + S_M \ln(t/\tau)$, where $M_0$ is the initial magnetization at the starting measurement time, $\tau$ is a characteristic time constant, and $S_M$ is the magnetic viscosity.[44] Magnetic viscosity is defined as the time dependence of magnetization in a constant magnetic field and temperature, which depends on the *spontaneous magnetization*, temperature, and "activation volume" of the system.[45] Experimentally, $S_M$ is positive for an increasing magnetization and negative for a decreasing one.[43]

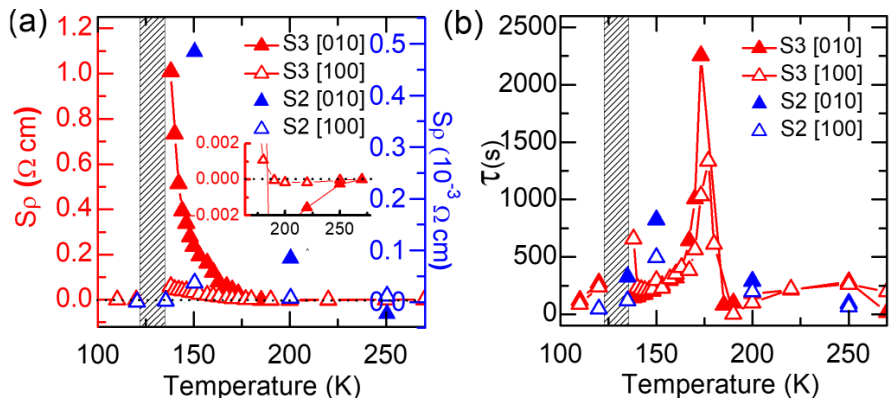

**Fig. 8.** Temperature dependence of the resistive viscosity $S_\rho$ (a) and relaxation time $\tau$ (b) obtained by fitting the resistivity relaxation curves to Eq. (1). For clarity, the resulting $S_\rho$ for S3 and S2 are plotted with the left and right y-axis respectively. The inset in (a) shows a close-up of $S_\rho$ for S3, in a region where $S_\rho$ changes sign. No reliable fitting can be obtained in the hatched area near the phase boundary between the FMM-dominated and AFI-dominated states for S3, where the transport is dictated by the percolation of the FMM phase.

The complex relaxation behaviors in perovskite manganites have been investigated mostly by phenomenological approaches involving characteristic relaxation time and magnetic or resistive viscosity. Close correlations between relaxations of magnetization and resistivity have been observed and analyzed.[29,46] Specifically, in a description of the dynamic phase separation in LPCMO proposed by Ghivelder, Parisi and Quintero,[44,46] the dynamics is dictated by the phase competition in the EPS state, rather than the magnetic interactions between individual magnetic moments as in spin glasses. Relative volume fraction of the coexisting phases in the EPS state changes with time, resulting in the relaxations of magnetization and resistivity. A large value of $S_M$ suggests a relatively large fraction of the system is involved in the dynamics.[47] Along this line, quantitatively, most of our resistivity relaxation curves exhibit logarithmic time dependence, which can be described by the modified equation,[47,48]

$$\rho(T,t) = \rho_0 + S_\rho(T)\ln\left(\frac{t}{\tau}+1\right), \qquad (1)$$

where $\rho_0$ is the initial resistivity, $\tau$ is the characteristic relaxation time, $S_\rho$ is the resistive viscosity. Analogous to the role of $S_M$ in the magnetization dynamics, the resulting $S_\rho$ can be regarded as a measure of the relative fraction of the system switching to the insulating phase; a positive $S_\rho$ implies a larger fraction of the system tends to switch from the FMM phase to the AFI phase than the reverse within the duration of the measurements. The $S_\rho$ and $\tau$ obtained from the fittings are plotted in Fig. 8.

Focusing on $S_\rho$ for S3, no obvious relaxation exists in the FMM-dominated state below 120 K, as $S_\rho$ is essentially zero in both directions. Entering the AFI-dominated state above 138 K, $S_\rho$ along [100] and [010] directions jump to finite positive values, indicating net isothermal fluctuations of relatively large fraction of the material from the FMM phase into the AFI phase over the duration of the relaxation measurement. With increasing temperature, $S_\rho$ decreases rapidly with increasing temperature, indicating the proliferation of the AFI phase and diminished relaxation from the FMM phase to the AFI phase. $S_\rho$ changes sign when the temperature reaches 183 K for [010] and 187 K for [100]. The negative $S_\rho$ above these temperatures indicate that the relaxation now favors a net gain of the FMM phase in the system at the expense of the AFI phase. It is worth noting that in the narrow temperature window between 130 K and 135 K, where the variation of the electrical conductivity is dictated by the *percolation* of the FMM phase rather than direct contribution from its volume fraction, Eq. (1) fails to describe the measured resistivity relaxation. Upon reaching 270 K, $S_\rho$ in both directions revert to zero, as expected. The relaxation times, $\tau$, for S3 in the two directions have similar temperature dependence. At the majority of measurement temperatures, $\tau$ has a magnitude of the order of $10^2$ s; however, $\tau$ for [010] and [100] both show clear indication of divergence at the temperatures near the equilibrium state where $S_\rho$ changes sign. The resistivity relaxation measurements and the fitting analyses therefore provide additional insights and a more quantitative picture of the evolution of the EPS phase dynamics with temperature, which are fully consistent with the static signatures of the EPS.

## IV. DISCUSSION

By adjusting the post-growth annealing time, anisotropic EPS in the optimally doped LCMO films has been engendered and controlled via the elastic strain arising from lattice mismatch between the substrate and epitaxial thin film. As the three films were grown under the same process to the same thickness, the post-growth annealing time is the only variable controlling the emergence of the anisotropic EPS. The discussion on the origin of anisotropic EPS will focus on the following two perspectives. First, the enhanced distortion and rotation/tilt of the $MnO_6$ octahedra under the promoted anisotropic strain give rise to the localization of the itinerant $e_g$ electrons, the onset of AFI phase, and the resultant EPS. Second, possible reasons for the self-assembled micrometer scale EPS of coexisting FMM and AFI phases will be discussed.

Emergent AFI phase and EPS are expected from the enhanced tilt and deformation of the MnO$_6$ octahedra, mediated by the specific anisotropic strain resulting from the post-growth annealing. Distinct from LCMO films on SrTiO$_3$ (STO) and LaAlO$_3$ (LAO) substrates, the coherent growth of LCMO films on (001) NGO substrates develop an anisotropic strain of *opposite signs* along the two orthogonal in-plane directions. STO and LAO substrates have been utilized extensively to induce biaxial isotropic strains (tensile or compressive) in LCMO thin films.[22,49] However, the large lattice mismatch between LCMO and STO and LAO substrates of cubic or pseudocubic structure adversely affects the coherency of the strained states, especially for relatively thick films.[22,50] While NGO with the same orthorhombic *Pbnm* symmetry as LCMO, in-plane anisotropic strain can be induced by the orthorhombic mismatch between NGO and LCMO. Moreover, with the small lattice and symmetry mismatches, the LCMO films can have better substrate-film coherency to form a strain state.[49,51] To optimize the crystallinity, *ex situ* post-growth annealing of the LCMO/NGO thin films not only improves the epitaxial quality and minimizes oxygen deficiency, but also enhances the substrate coherency.[51]

The as-grown S1 exhibits bulk-like transport properties since it is almost strain-free and retains its bulk orthorhombic symmetry. The prolonged post-growth annealing in S2 and S3 triggers and promotes the formation of AFI phase in the FMM ground state, as manifested in Fig. 2 and Fig. 4. Assuming LCMO film is perfectly locked on (001) NGO substrate, the in-plane lattice constant *a* along [100] should shrink while *b* along [010] needs to elongate, and the rotation about [001] and tilt about [100] should be further enhanced. Two cooperative octahedral responses for accommodating the anisotropic strain have been proposed.[51,52] The first originates from the robust octahedral tilt of the NGO substrate; a combination of the breathing mode (the length of Mn-O bonds increases) in JT distortion and a further tilting in LCMO film meet the requirement of the orthorhombic distortion. The other distortion mode is a further octahedral rotation about [001], together with the Q$_2$ mode JT distortion (one in-plane Mn-O bond stretches in one direction and another compresses in the orthogonal direction) in the *ab*-plane. The rotation and tilting in both modes make buckling of the Mn-O-Mn bond angle deviate from 180°, which significantly reduces the hopping probability of the itinerate electron in the double exchange scenario due to the reduction of the overlap of the Mn *3d* and O *2p* orbitals.[53] At low temperatures, the concurrent JT distortions further split the degeneracy between the e$_g$ and t$_{2g}$ orbitals and favor the AFI charge/orbital ordered state.[6]

The induced EPS also exhibit anisotropy indicated by static magnetoresistance measurements. The pronounced anisotropy in the electrical transport coincides with the appearance of the EPS and the strengthening of the anisotropic strain. These observations can be well accounted for in a picture of anisotropic percolation of the FMM phase in the EPS state. In other words, the FMM entities in the background of the predominant AFI phase have a well-defined orientational preference along [100] suggested by the smaller resistivity along [100]. The resistivity anisotropy RA gives a more quantitative measure of the thermal and field-induced anisotropic percolation.

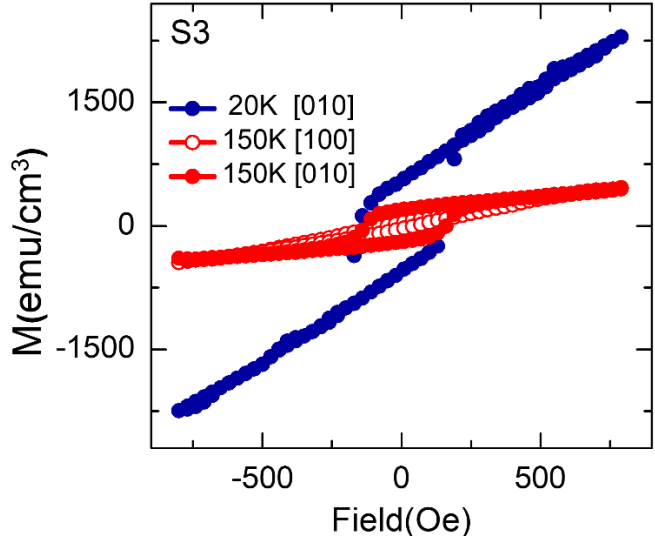

**Fig. 9.** Magnetization (M) curves at 150 K with a magnetic field applied along [100] and [010] direction, and at 20 K with the field along [010].

Although a comprehensive explanation for the anisotropic/patterned EPS is still lacking, we can obtain some insight by discussing/comparing our results with other experimental observations and theoretical simulations. For self-organized anisotropic electronic inhomogeneity at the micrometer scale, twin-domain/multidomain structure, modified orbital ordering (OO), and anisotropic energy landscape rising from elastic strain have been considered to be essential factors.[15,16,54,55]

First, magnetization measurement was performed to identify the magnetic easy axis in S3, and the results are shown in Fig. 9. After FC, M(H) was measured with the applied field parallel to [100] or [010] direction at 20 K and 150 K, corresponding to the FMM-dominated and AFI-dominated regimes in the EPS state respectively. First, the saturation magnetization at 20 K (in the FMM-dominated state) is significantly higher than that at 150 K (in the AFI-dominated state), which is due to the prevailing AFI phase at the expense of the FMM phase at 150 K. Moreover, at 150 K, M(H) exhibits distinct anisotropy between the [100] and [010] directions. Although the magnetization does not reach saturation in either direction, the larger coercive field and remnant magnetization in [010] direction than those in [100] direction identify [010] as the magnetic easy axis. However, the magnetotransport results clearly indicate that the FMM domains preferentially grow along [100] direction. A similar contradiction has been observed in LCMO/ (100) NGO thin films. A twin-domain structure has been speculated to be induced from a shear-mode deformation of the $MnO_6$ octahedra, which was used to explain the anisotropic EPS.[31] The $MnO_6$ octahedra could tilt about [100] in opposite directions and lead to twin domains in LCMO film. The incommensurate lattice between the twin domains disallows JT distortion and favors the formation of FMM phase along [100]. The large shear-mode strain and a concomitant twin structure were experimentally observed by Wakabayashi *et al.* in NSMO/(011) STO films.[15] Through the structural analysis via X-ray diffraction at low and high temperatures (10 K and 200 K), a clear large scale phase coexistence of the undistorted high-temperature phase and the distorted low-temperature phase was revealed. The remnant high-temperature phase originates from the large stress at *martensitic* twin-domain boundaries, which disallows the JT distortion and suppresses the orbital ordering.

The anisotropical suppression of OO by a specific strain state has been found to account for the large scale in-plane anisotropic EPS in the strained NSMO/(110) STO films.[55] Discernible "nematic" domains statistically favor alignment along [1$\bar{1}$0]. It is believed that the transition between OO and non-OO states requires JT-like distortion in the OO planes. The OO planes are parallel to the (100) or (010) planes, lying out of the film surface. Although the in-plane lattice constants are fixed to those of the substrate and disallowed for the JT-like distortion, the lattice relaxation along [1$\bar{1}$0] favors JT-like distortion and formation of OO domains, resulting in the anisotropic growth of the FMM domains.[54] In our films, further simultaneous studies of the structure and electronic states are needed to ascertain the correlations between specific structural variations with the onset of anisotropic EPS under the induced anisotropic strain.

On the other hand, primarily based on the structural aspect, model simulations for manganites with strong electron-lattice coupling have demonstrated that the strain-induced atomic scale lattice distortions can lead to self-organized electronic inhomogeneities at both nanoscale and microscale.[16] The strain-induced distortions are considered to be a combination of long range intracell modes and short range intercell modes.[16,17,19] Concurrent intercell and intracell modes, and constraints among them, drive the formation of rich energy landscapes energetically favoring different phase configurations. The interaction and coupling between the distortion modes can become highly anisotropic, and generate energy landscape anisotropically favoring certain phases.[18] The deformation (shrinkage of lattice constant $a$ and elongation of $b$), the tilt and the rotation of the $MnO_6$ octahedra might modify the overall intrinsic energy landscape of system, and favor an orientation-preferred growth of the FMM phase in the EPS state.

## V. CONCLUSIONS

We have experimentally demonstrated that anisotropic microscale EPS can be induced by post-growth annealing in optimally doped LCMO/(001)NGO thin films via static magnetotransport and dynamic relaxation measurements along the two orthogonal in-plane directions. In the EPS state, both the temperature and magnetic field dependent static transport properties exhibit apparent and consistent resistivity anisotropy. The resistivity along the tensile-strained [010] becomes significantly larger than that along the compressive-strained [100], suggesting the FMM entities in the EPS state have an elongated morphology along [100].

Resistivity relaxation measurements after field-cooling provide a more quantitative measure of the dynamic competition between the coexisting AFI and FMM phases in the EPS state. Analyses based on a phenomenological model reveal that at the boundary between the FMM-dominated and AFI-dominated states, the dynamics favor a net gain of the AFI phase at the expense of the FMM phase. The isothermal dynamic growth of the AFI phase gradually diminishes with increasing temperature in the AFI-dominated state. This evolution of dynamics with respect to temperature correlates closely with the signatures of EPS in the static measurements. The resistivity anisotropy in the EPS state shows consistent enhancement with annealing time, which suggests that the strain-mediated deformations of the MnO$_6$ octahedra are the microscopic origin for the emergence of the anisotropic EPS, as the post-growth annealing at a sufficiently high temperature gradually improves the film-substrate coherency and enhances the anisotropic strain.

Due to the particular anisotropic strain of *opposite signs* along the two orthogonal in-plane directions, we surmise that the enhanced deformations of MnO$_6$ octahedra not only trigger the formation of the AFI phase in the FMM ground state, but also plays a critical role in the anisotropic distribution of FMM entities in the EPS state. Possible mechanisms have been discussed by comparing our results with the large scale anisotropic EPS in other manganites. The approach of engineering the anisotropic EPS and its dynamic properties by tuning the anisotropic strain via post-growth annealing may be applied to other strongly correlated oxide films.

## ACKNOWLEDGMENTS

The authors thank Stephan von Molnar for helpful discussions. This work was supported by NSF grant DMR-1308613 (P.X.), the National Natural Science Foundation of China (Grant 11474263 and 11674312), the National Basic Research Program of China (Grant 2016YFA0401003), and Hefei Science Center CAS (Grant 2016HSC-IU006).